\documentclass[prb,preprintnumbers,amsmath,amssymb,floatfix,aps,twocolumn,superscriptaddress]{revtex4} 
\usepackage{graphicx}
\usepackage[utf8]{inputenc} 

\usepackage{xcolor}

\usepackage{hhline}
\usepackage{hyperref}

\usepackage{mathabx}

\newcommand\doilink[1]{\href{http://dx.doi.org/#1}{#1}}
\newcommand\doi[1]{DOI:\doilink{#1}}

\renewcommand*\url[1]{\href{#1}{\texttt{#1}}}

\begin{document}

\title{Strain-induced phase transition in CrI$_{3}$ bilayers}

\author{Andrea Leon}
\email{andrea.leon@postgrado.usm.cl}
\affiliation{Max Planck Institute for Chemical Physics of 
Solids, Dresden, Germany.}

\author{J. W. Gonz\'alez}
\email{jhon.gonzalez@usm.cl}
\affiliation{Departamento de F\'{i}sica, Universidad 
T\'{e}cnica Federico Santa Mar\'{i}a, Casilla Postal 
110V, Valpara\'{i}so, Chile.}

\author{J. Mej\'ia-L\'opez}
\affiliation{Centro de Investigaci\'on en Nanotecnolog\'ia y Materiales Avanzados CIEN-UC, Facultad de F\'isica, Pontificia Universidad Cat\'olica de Chile. CEDENNA, Santiago.}

\author{F. Crasto de Lima}
\affiliation{Instituto de F\'isica, Universidade Federal 
de Uberl\^andia, Uberl\^andia, MG,  Brazil.}

\author{E. Su\'arez Morell}
\email{eric.suarez@usm.cl}
\affiliation{Departamento de F\'{i}sica, Universidad 
T\'{e}cnica Federico Santa Mar\'{i}a,
Casilla Postal 110V, Valpara\'{i}so, Chile.}

\date{\today}

\begin{abstract}
A monolayer of CrI$_3$ is a two-dimensional crystal in its equilibrium configuration is a ferromagnetic semiconductor. In contrast, two coupled layers can be ferromagnetic, or antiferromagnetic depending on the stacking. 
We study the magnetic phase diagram upon the strain of the antiferromagnetically coupled bilayer with C2/m symmetry. We found that strain can be an efficient tool to tune the magnetic phase of the structure. A tensile strain stabilizes the antiferromagnetic phase, while a compressive strain turns the system ferromagnetic. We associate that behavior to the relative displacement between layers induced by the strain. We also study the evolution of the magnetic anisotropy, the magnetic exchange coupling, and how the Curie temperature is affected by the strain.
\end{abstract}


\maketitle

\section{\label{sec:intro} Introduction}

The synthesis and manufacture of two-dimensional materials (2D) has  
revolutionized solid-state physics, dozens of monolayer materials 
have been synthesized and measured their properties while many others 
have been predicted theoretically allowing us to think in 
\textit{á la carte} 
design of devices\cite{Novoselov2016,PhysRevX.3.031002}. These 2D materials have 
already broken some record properties and we now have at hand metals, semiconductors, 
topological insulators and trivial insulators that might be used in practical devices\cite{li2019intrinsic,wang2018very}. 
A 2D material does not necessarily imply a single atomic layer but a less strict 
definition that implies a different behavior than the bulk. In fact, assembling 
layers of these materials in different ways has led to the discovery of 
correlation phenomena like superconductivity\cite{Cao2018}. 

Although many of the properties that arise in 2D materials have already been explored, the intrinsic magnetic ordering was experimentally found, first in a monolayer of iron phosphorus trisulfide\cite{Lee2016,Wang2016} FePS$_3$, in a chromium triiodide monolayer (CrI$_3$) and in a bilayer CrGeTe$_3$ structures\cite{Huang2017,Gong2017}. The first one has intrinsic antiferromagnetic order while the other has ferromagnetic ordering. 
Moreover, a van der Waals (vdW) layered  Fe$_{3}$GeTe$_2$ crystal remains ferromagnetic when 
thinned down to a monolayer\cite{Fei2018}.  The exploration of 2D magnetic phases and controlled 
transitions can lead to envision novel devices\cite{gong2019two,Geim2013}.

The chromium  halides (CrX$_3$, X=Cl, Br, I) belong to the family 
of vdW layered materials, with an hexagonal crystal structure 
within the layer and with layers weakly bonded by vdW forces. 
Its magnetic ordering temperature increases with
the size of the halogen atom from Cl to I. The three structures have a ferromagnetic (FM) 
order within the layers and an antiferromagnetic (AF) coupling between 
layers\cite{soriano2019interplay,wang2018very}. 
The superexchange mechanism which favors FM ordering is more important than the direct 
interaction between Cr atoms\cite{Sivadas2018}. In CrCl$_3$ the spins are aligned parallel 
to the layers plane while in CrBr$_3$ and CrI$_3$ the direction of the 
spins is perpendicular to the plane\cite{webster2018strain}. 
The spin-orbit coupling of the halides 
increases with the size of the atom and this has been identified as the origin 
of the magnetic anisotropy\cite{lado2017}.    
A bulk crystalline CrI$_3$ exhibits Ising ferromagnetism below a Curie 
temperature\cite{liu2019thickness,Huang2017} (T$_C$) of 61 K, it was not surprising to find a 
ferromagnetic ordering in a monolayer of this material. However, the bilayer structure displays 
an antiferromagnetic behavior\cite{Huang2017}. In addition, it is possible to tune the magnetic 
phase  with an external electric fields\cite{Huang2018,Jiang2018,morell_2019}, with their evident 
implications in designing new electronic components.

\begin{figure}[hb]
\centering
\includegraphics[clip,width=0.5\textwidth,angle=0]{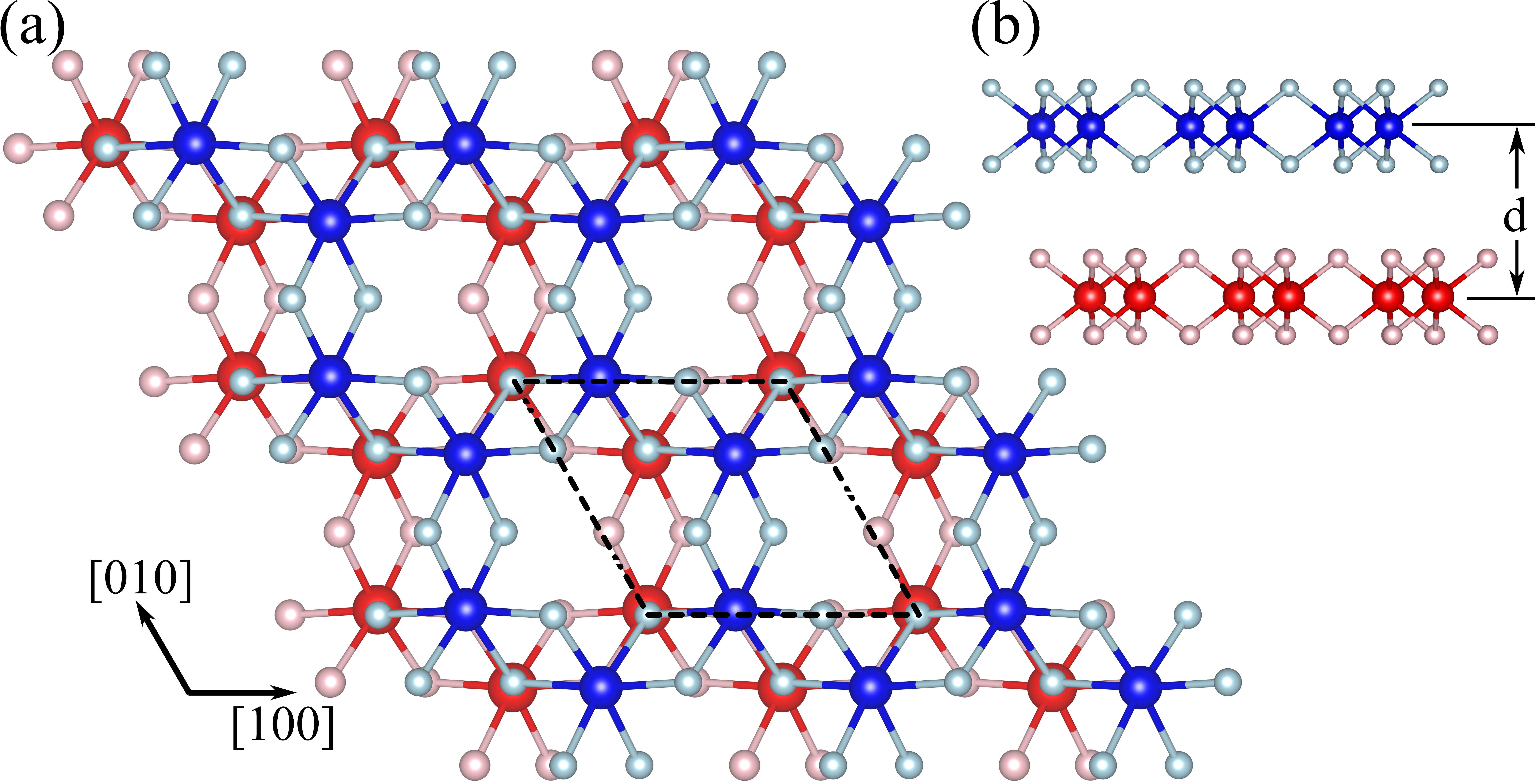} 
\caption{Scheme of the CrI$_3$ bilayer in the C2/m symmetry. Top (a) and side view (b). For the sake of clarity, we have used bluish tones for the top 
layer and reddish tones for the bottom layer. }
\label{Fig:Esq}
\end{figure}

A systematic study of the bulk CrI$_3$ crystal reveals that there are 
two main structural phases, one is a monoclinic arrangement with a 
C2/m space group symmetry\cite{jiang2019stacking,jang2019microscopic}  
(see fig. \ref{Fig:Esq}), which 
is the most stable configuration when the temperature is higher than 
$\sim 200$ K,\cite{mcguire2015coupling} below that temperature  a phase transition 
occurs to a rhombohedral stacking, space group R$\bar{3}$. 
Density Functional Theory (DFT)  studies have revealed that bilayer structures 
with the two mentioned symmetries have different magnetic ordering, 
the C2/m bilayer structure is antiferromagnetic while the R$\bar{3}$ bilayer structure 
is ferromagnetic\cite{soriano2019interplay,morell_2019}.
This fact apparently contradicts the measured magnetic phase of the bilayer 
structure\cite{jiang2019stacking}, taking into account that the experimental measurements were done at 
temperatures well below 200 K. However it is noteworthy to mention that the 
structures are fabricated at high temperatures 
and then encapsulated, which might favors the appearance of the C2/m stacking  at
low temperatures\cite{wang2018very,Huang2017}, CrI$_3$ flakes are usually sandwiched between graphite layers to prevent reaction with oxygen and moisture\cite{Huang2017}.

The magnetic phase of a single layer of CrI$_3$ can be controlled by strain, 
where a 6\% compressive strain has shown to induce an FM to AF 
transition\cite{webster2018strain}.  The transition can be understood in terms 
of a competition between the AF Cr-Cr direct exchange interaction and the FM superexchange interaction.  
A compressing  strain will bring the two chromium atoms closer and the system becomes 
antiferromagnetic\cite{mokrousov2007magnetic}. 
A hypothetical structure with only the Cr atoms is AF, however the superexchange interaction between the Cr atoms mediated by the iodine atoms gives rise to the Cr-Cr FM coupling\cite{lado2017}. It is possible then to visualize the magnetic behavior of the monolayer as a competition between the two mechanisms mentioned above, for the non-strained system, the FM interaction prevails.
It is worth mentioning, however, that we have found some discrepancies that we discuss later on about whether the transition from FM to AF occurs with compressive or tensile strain. 

It has recently been studied by two independent groups that a bilayer CrI$_3$ system undergoes a magnetic phase transition when hydrostatic pressure is applied. The system changes from AF to FM and the transition may be irreversible\cite{Song2019,Li2019}. Similar studies were carried out in multilayers of CrCl$_3$, we show that our theoretical results agree with these results\cite{Azkar2019,Wang2019}.

In this work, we focus on the strain-induced effects in the electronic
and magnetic properties of CrI$_3$ bilayer in the C2/m phase, we will 
refer it as the HT (high temperature) stacking. This stacking features FM 
coupling between chromium atoms within the layers and AF coupling 
between chromium atoms of different layers. We found that for a small 
compressing  strain around -2\%, the system undergoes an AF to FM coupling between layers and that change, with an applied strain, 
can be related to an induced shift between layers. We show that a slight relative  shift between layers changes the magnetic ordering of the structure.
This result goes in the same direction that the conclusions of the two mentioned experimental articles\cite{Li2019,Song2019}.
Therefore, the strain might be an efficient tool to control the magnetic properties of bilayer CrI$_3$.  The tensile strain enhances the stability of the AF ordering between layers, while the compressive strain changes the structure to FM ordering. If we keep the strain, the created high potential barrier  makes difficult to revert the system to the original AF state.

We also study the evolution with the applied strain of the magnetic exchange 
coupling between Cr atoms of the same and different layers, how  
the band structure changes with strain and the energy needed to change the magnetic 
ordering of one layer given a specific order in the other, and finally, we show 
how the Curie temperature of the bilayer structure is affected by the strain.

\begin{figure}[!h]
\centering
\hspace{-5mm} \includegraphics[clip,width=0.5\textwidth,angle=0]{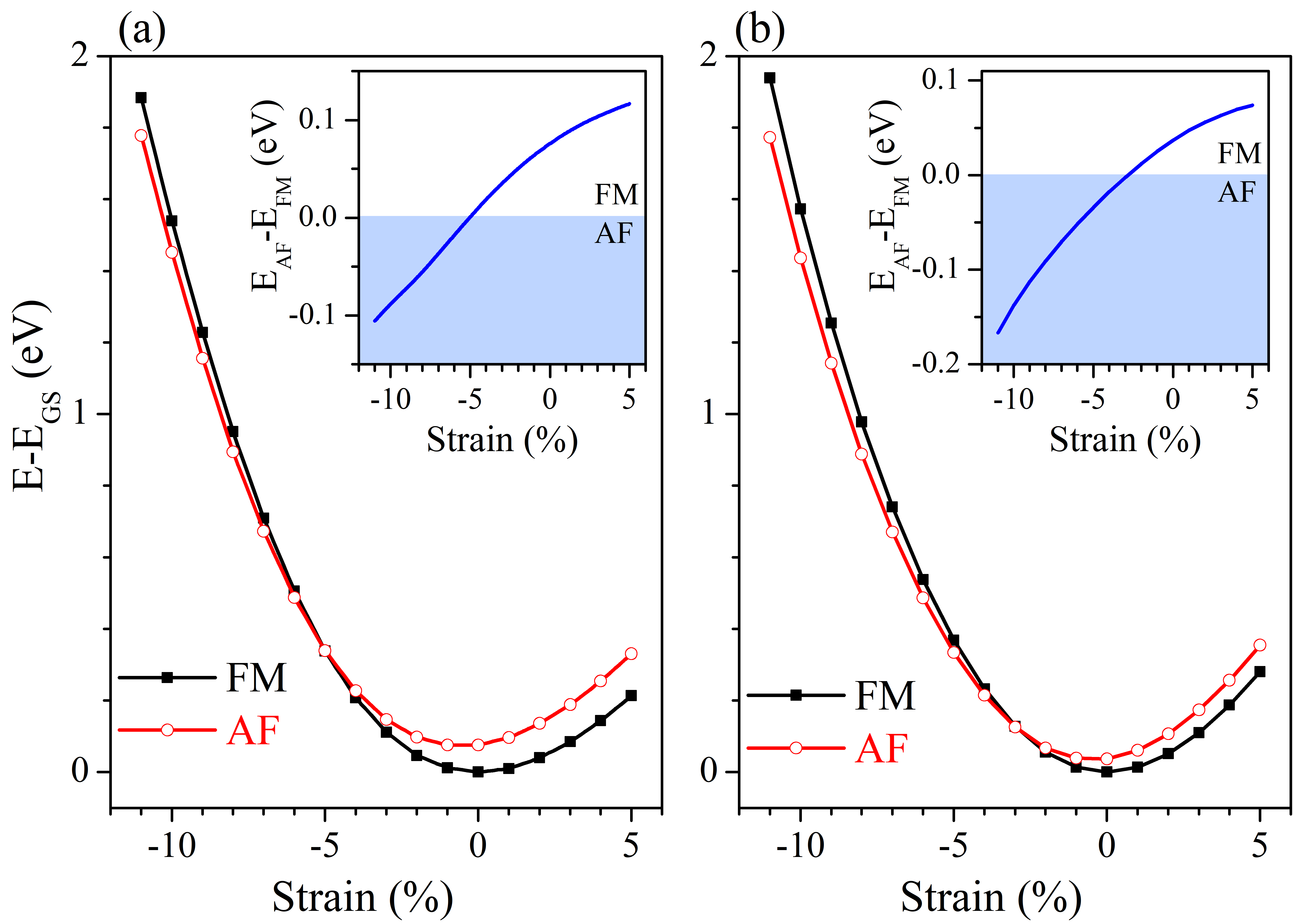} 
\caption{Total energy as a function of strain for the two magnetic ordering 
in a  monolayer of CrI$_3$ calculated with PBE in (a) and LSDA+SOC in (b). 
The reference energy E$_{\mathrm{GS}}$ is the energy of the FM case with 0\% strain.
The inset shows the energy difference between them: E$_{AF}$-E$_{FM}$. }
\label{Fig:Energy_mono}
\end{figure}

\section{\label{sec:method} Methodology}

In fig. \ref{Fig:Esq}, we show the unit cell of the HT bilayer structure; 
the system has a C2/m group symmetry, with two chromium atoms per 
layer forming a honeycomb lattice, each Cr atom is surrounded by six iodine 
atoms conforming a distorted edge-sharing octahedron.  
The equilibrium configurations are found relaxing the atomic positions 
and unit cell vectors using high convergence criteria. Without strain, 
the value of the lattice constant depends slightly on the magnetic phase, 
however, to reduce the number of free parameters, we fix, for all configurations, 
the lattice constant of the ground state. Therefore, for the monolayer we 
use the lattice vectors of the FM configuration, and for the bilayer 
we use the vectors from the phase where the coupling between layers is antiferromagnetic.
The strain is defined as the scaling of the in-plane lattice vectors 
--directions [100] and [010]--. Positive values correspond to a stretch, 
and negative values correspond to a compression of the structure\cite{gonzalez2019highly,webster2018strain}.

Our calculations were done with the 
OpenMx-code\cite{yoon2018reliability,ozaki2003variationally,han2007magnetic}, and  
validated with VASP\cite{VASP} calculations.
The calculations for the bilayer structures always include van der Waals 
interactions and a Hubbard repulsion U-term. We employed a value of U = 3 eV, but we 
 tested that the results are robust, for values of U between 2-4 eV, the transition is always observed, 
with small differences in the critical phase-change strain value.
In section \ref{compa}, we provide the structural parameters and compare the local and non-local models for the dispersion interactions.

In all our collinear calculations, we have used a Perdew-Burke-Ernzerhof (PBE) 
exchange\cite{PBE},  and in noncollinear 
calculations we used the local spin-density approximation LSDA to 
avoid gradient divergence problems. The interstitial region increases with 
tensile strain and it leads to convergence issues for the PBE noncollinear calculations\cite{blonski2010magnetocrystalline,peters2016correlation,gonzalez2017complex}.
Detailed information about DFT parameters is provided in the supplementary material.

\section{\label{sec:result} Results and Discussion}
\subsection{Stability Order}
\subsubsection{Monolayer}
At the moment, the physical properties of mono- and bi-layers of CrI$_3$ are in 
constant  development and there are still some disagreements. 
In particular, we want to clarify a discrepancy in the strain-driven transition 
from FM to AF state of the monolayer. 
On the one hand, using noncollinear calculations including a spin-orbit interaction 
Webster et al. \cite{webster2018strain} show that the transition occurs for negative 
values of the strain (compression). On the other hand, using collinear calculations 
Wu et al. \cite{wu2019strain} claimed that the transition occurs for positive strain 
values (expansion).
Our results using PBE and LSDA with and without spin-orbit coupling (SOC) 
always show that phase transition occurs for biaxial compression. 
There are small discrepancies between the calculations with PBE, LSDA, and LSDA+SOC. 
For instance, in the collinear calculation within PBE functional the FM/AF transition 
occurs at -5\% strain compression while for the calculation within LSDA and LSDA+SOC 
functionals the transition occurs at a compression of -3\%. In fig. \ref{Fig:Energy_mono}, 
we show how the total energy changes for the two magnetic 
ordering FM and AF and for PBE and LSDA+SOC as a function of
the strain, a transition from FM to AF occurs for a compressing  
strain. This is consistent with the increase of the AF direct
exchange coupling between the two Cr atoms as the atoms get closer\cite{mokrousov2007magnetic}.   

The strain induces small displacements in the atomic positions 
modifying the angle of the intralayer Cr-I-Cr bond responsible 
for the FM superexchange mechanism\cite{jiang2019stacking}.  
For the FM layers, we obtain that the Cr-I-Cr angle changes 
monotonically with the strain, obtaining that for 0\% strain is 
91.2$^\circ$, for -5\% strain is 87.4$^\circ$ and for 5\% is 
96.2$^\circ$. 

\begin{figure}[!ht]
\centering
\includegraphics[clip,width=0.48\textwidth,angle=0]{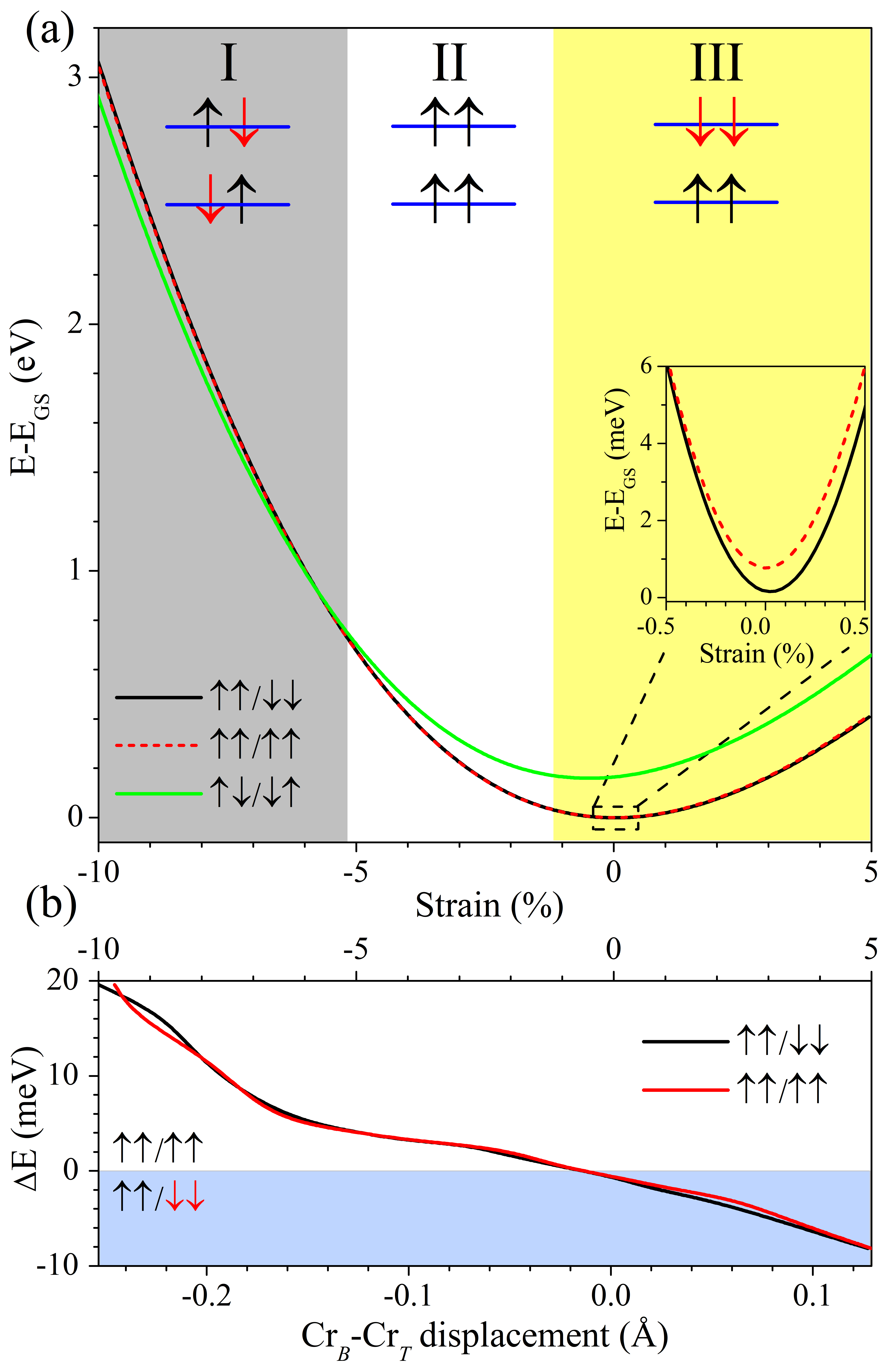}
\caption{(a) Total energy per unit cell for three different magnetic 
ordering for two layers of CrI$_3$ with C2/m stacking as a 
function of the strain. In each of the three shadowed regions 
a different magnetic phase is the most stable. The inset shows 
the behavior around 0\% strain of the two lowest phases 
(AF $\upuparrows/\downdownarrows$ and FM $\upuparrows/\upuparrows$).
The reference energy E$_{\mathrm{GS}}$ is the total energy of the $\upuparrows/\downdownarrows$ configuration at 0\% strain.
(b) Change in total energy 
($\Delta E = E_{\upuparrows/\downdownarrows}-E_{\upuparrows/\upuparrows}$) 
as a function of the strain-induced displacement between second 
interlayer Cr neighbors. Cr$_{B/T}$ labels Cr atoms of bottom/top layer. As a reference, at 0\% strain the in-plane distance Cr$_B$-Cr$_T$ is 2.27 \AA{}. 
For the sake of clarity, we have included an additional top x-axis with the applied strain. 
}
\label{Fig:Energy}
\end{figure}

\subsubsection{Bilayer}

We proceed now with the bilayer HT structure, the non-strained system is AF ($\upuparrows/\downdownarrows$).
The equilibrium lattice constant obtained with our calculations with both PBE and LSDA  functionals are 6.9 \AA{}  
and  6.7 \AA{} respectively.
The lattice constant identified using X-ray diffraction and absorption spectroscopy techniques in 
bulk/few-layers\cite{frisk2018magnetic,Huang2017,mcguire2015coupling} CrI$_3$ is 
6.87 \AA{}. As expected, compared with experimental data, the LSDA underestimates the 
lattice constant, while PBE overestimates it. in what follows all our results are using the PBE functional unless otherwise stated. 

The distance between layers changes with the strain, the distance without strain is 6.6 \AA{}, see fig. \ref{Fig:Esq}; 
it diminishes with a tensile strain, for an expansion of 5\%, we obtain 
a distance of 6.4 \AA{}, and for a compression of -5\%, the distance 
increases up to 6.8 \AA{}. 
The above result can be understood as the rearrangement of the atomic positions induced by strain. As the system stretches and the separation between atoms of the same layer increases, allowing the minimization of energy through a more compact configuration in the direction perpendicular to the atomic plane\cite{cortes2018stacking}.

In general, for the bilayer structure, we must consider two types of 
magnetic orders, one referring to the interaction within the same 
layer and the other referring to the interaction between layers. 
If we consider in all cases out of plane magnetization there are 
three non-degenerated states: I) AF in each layer 
$\updownarrows/\updownarrows$ , II) FM in both layers 
$\upuparrows/\upuparrows$ and III) FM in each layer and 
AF as a whole $\upuparrows/\downdownarrows$. In fig. \ref{Fig:Energy}a, 
we plot how the total energy for each of these three states 
changes with the strain, we shadowed each of the aforementioned 
regions in the figure. We learned from the monolayer that for compressing strain below 
-5\%, the interaction within the layers is antiferromagnetic. Such behavior appears in 
the bilayer, where below -5\% the curve belonging to the configuration I becomes 
energetically more favorable. In region II for compressing  strain between 
-1.2\% and -5\% the system is FM and in region III that 
corresponds to strain values above -1.2\%  is characterized 
by an antiferromagnetic coupling between layers. The energy difference between the configurations 
$\upuparrows/\upuparrows$ and $\upuparrows/\downdownarrows$ is 0.5 meV without strain.
The transition between the regions I and II is due to the same mechanism we discussed for the FM to AF transition in the monolayer, where compression favors the AF coupling between Cr atoms of the same layer.

The magnetic ordering of two coupled monolayers of CrI$_3$ depends on the stacking, it was shown previously that a relative displacement of the 
layers impacts the magnetic ordering\cite{Sivadas2018}. We show that the origin of the magnetic transition with the compressive strain is related with a change in 
the in-plane distance between Cr atoms of different layers. 
In fig. \ref{Fig:Energy}(b), we correlate the horizontal relative strain-induced movement of the Cr atoms of different layers and the magnetic phase of the system. For the non-strained system, we found that a small shift in the horizontal plane has a more significant impact on the magnetic phase than a vertical compression between layers. 

To understand how the displacements between layers change the magnetic phase --without strain--.  
In the supplemental material (section \ref{sec:Disp} and fig. \ref{Fig:dxyz_bi}a), we show that the magnetic phase changes with the horizontal displacement between layers along the direction [110], as we found under biaxial strain. And in turn, no changes with the magnetic phase are found when layers are vertically compressed; the layers need to be decoupled to obtain a transition to the FM state, which is more difficult to achieve experimentally. 

An alternative method, recently suggested in two experimental articles \cite{Li2019,Song2019}, for achieving a magnetic phase transition would be a  crystallographic phase change from HT to LT driven by external pressure, this is consistent with the mechanism we propose. Furthermore, the authors observed an irreversible magnetic phase change, probably since at low temperatures the most stable structural phase is the rhombohedral (LT) which has been predicted to be FM\cite{morell_2019, Sivadas2018}. In the LT structure the difference between the FM and AF phases is 4 meV/Cr atom,\cite{morell_2019} for a magnetic transition, becomes a large barrier to overcome.  
%

The band structure for the FM and AF phases between -6 eV and 1 eV are dominated by the $d$-Cr orbitals and the $p$-I orbitals and  states appearing at 
energies above 3 eV are mostly $d$-Cr orbitals. The bands around 1 eV of the FM phase ($\upuparrows/\upuparrows$) correspond exclusively to spin up components and around 3 eV they are spin down. 
This composition of the band structure, the mixing of $d$-Cr and $p$-I orbitals\cite{liu2016exfoliating,wu2019strain}, favors the supersuperexchange interaction 
between Cr-I$\cdot \cdot \cdot$I-Cr of different layers\cite{Sivadas2018}.

The band structure of the bilayer structures are slightly modified 
upon strain. The modifications induced by the strain are more 
evident in the valence bands around the $\Gamma$-point, as already 
observed in other semiconductors subjected to strain\cite{gonzalez2019highly,scalise2012strain}. When the system is stretched, the band structure becomes flatter, and by compressing the structure, the bands become wider. However, we do not observe substantial modifications in the band structure nor band composition that could indicate changes in the magnetic interaction mechanisms --superexchange and supersuperexchange--.

\begin{figure}[!h]
\centering
\includegraphics[clip,width=0.48\textwidth,angle=0]{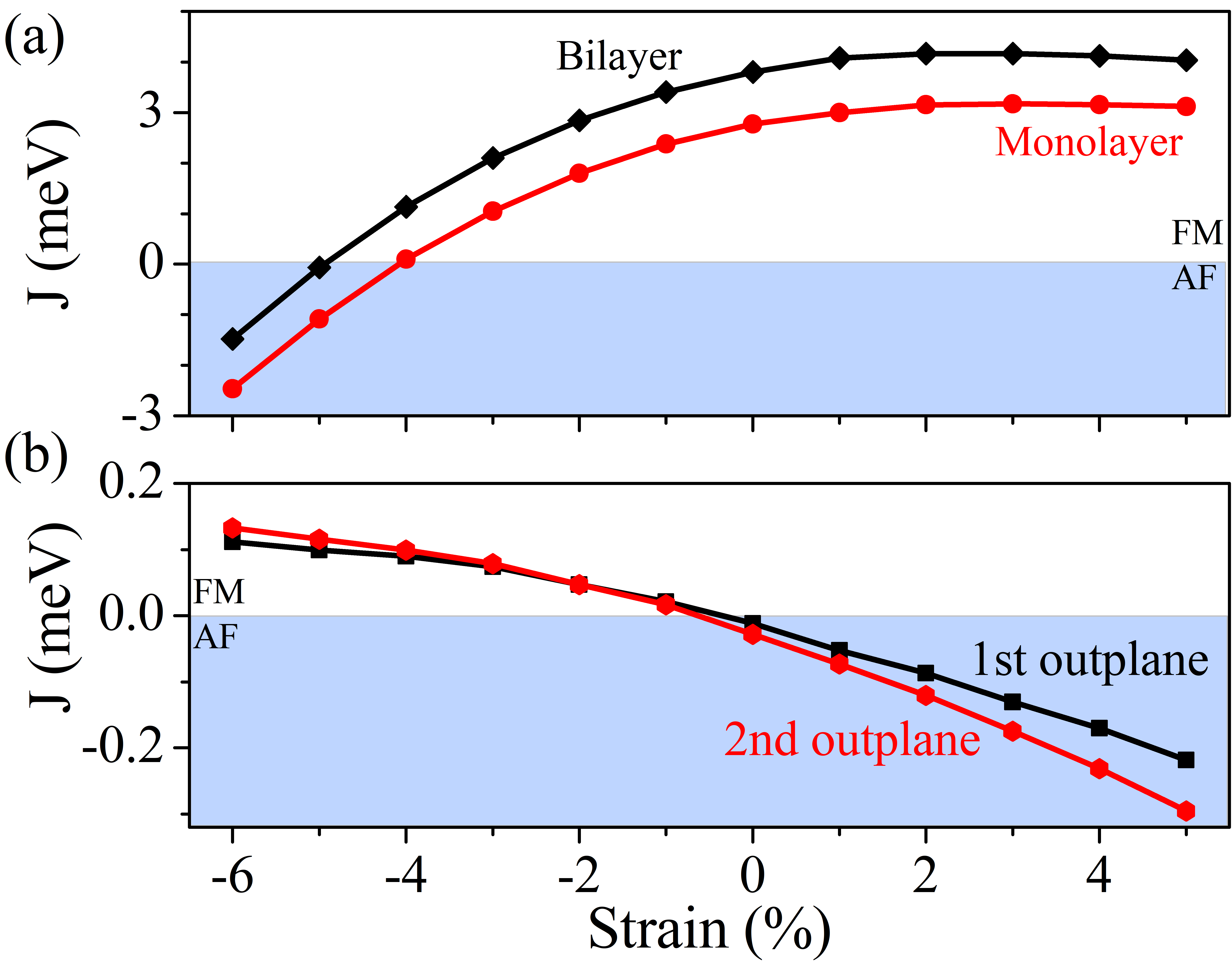} 
\caption{In-plane Cr-Cr exchange coupling J as a function of strain for CrI$_3$ monolayer and bilayer (top panel), and Interlayer Cr-Cr exchange coupling J as a function of the applied strain for the first and second Cr neighbor (bottom panel). }
\label{Fig:Jbi}
\end{figure}

\subsection{Magnetic Exchange Interaction J and Curie Temperature}
We calculated the exchange coupling parameters using the magnetic force theorem (MFT), based on a second order perturbation theory\cite{yoon2018reliability,jang2019microscopic,ozaki2003variationally} within the OpenMX package.
In monolayer CrI$_3$ the value of the exchange energy J remains almost constant for tensile strain while for compressive strain the value tends to diminish and for a strain lower than -4\%, J changes sign and the system becomes antiferromagnetic (fig. \ref{Fig:Jbi}a). This result is consistent with what we found earlier. A similar behavior is found in each layer of the bilayer structure although the value is slightly higher. We checked that this difference diminishes when we separate the layers, i.e. going to the decoupled monolayer limit.

In the bilayer structure we  consider the interaction between Cr atoms of different layers. In fig. \ref{Fig:Jbi}b we show how the coupling exchange for the first and second nearest interlayer neighbors behave under strain.
In both cases the value of J shows a change of sign for a critical strain of $\sim$1\%.

We now proceed with the calculated J values to obtain the thermodinamical properties of the system using the Ising model\cite{chen2019boosting,miao20182d,lu2019mechanical}. 
Herein, we use a Metropolis Monte Carlo simulations to numerically solve the 2D Ising model\cite{paez2007computational} and calculate the specific heat,  estimate the Curie temperature and its response upon strain in CrI$_{3}$ mono- and bi-layer. More information on the applied Ising model is included in the Supplemental Material.

We found a Curie temperature of T$^M_C$=44.4 K for the non-strained monolayer CrI$_3$. A tensile strain of +3\% slightly increases the T$^M_C$ reaching 51.4 K. We showed before that the exchange coupling J is more susceptible to compressive strain, therefore the Curie temperature and the specific heat changes significantly with compression, for instance, the Curie temperature decreases to a value of T$^M_C$=14.1 K for a compressing strain of -3\%.  
The values of J and T$_C$ found for monolayer and bilayer for 0\% strain agree with previous theoretical and experimental reports\cite{Huang2017,kim2019evolution}.

Although the values of in-plane J$_{Cr-Cr}$ for the bilayer is slightly larger than in the monolayer case the Curie temperature is slightly lower by $\sim 4$ K in the non-strain case.
For zero and positive strain, the bilayer Curie temperature T$^B_C$ 
decreases compared with its monolayer analogue T$^M_C$.  The discrepancy 
increases with the strain. For -3\% strain the Curie temperature increases  while for zero and positive strain, the T$^B_C$ 
decreases compared with the monolayer value.

The exchange interaction J and the Curie temperature are directly related\cite{soriano2019interplay}. In both cases: monolayer and bilayer structures, the value of the intralayer J diminishes with a compressive strain as it does the Curie temperature. 
However, the value of J in the bilayer structure includes two contributions, one intralayer and another one interlayer. The intralayer exchange rapidly decreases with the compressive strain --being slightly larger than in the monolayer case (fig. \ref{Fig:Jbi}(a))--. However, for a compressive strain above $\sim$ -1\%, the interlayer exchange interaction increases and changes sign (Fig. \ref{Fig:Jbi}(b)) that results in a faster decrease of the Curie Temperature of the monolayer structure. That explains the change in the order of the Curie temperature observed for compression strain, where the T$^M_C$ $<$ T$^B_C$.
For further information, we refer the reader to the section \ref{sec:Ising} and fig. \ref{Fig:isi_bi}.

\begin{figure}[!ht]
\centering
\includegraphics[clip,width=0.48\textwidth,angle=0]{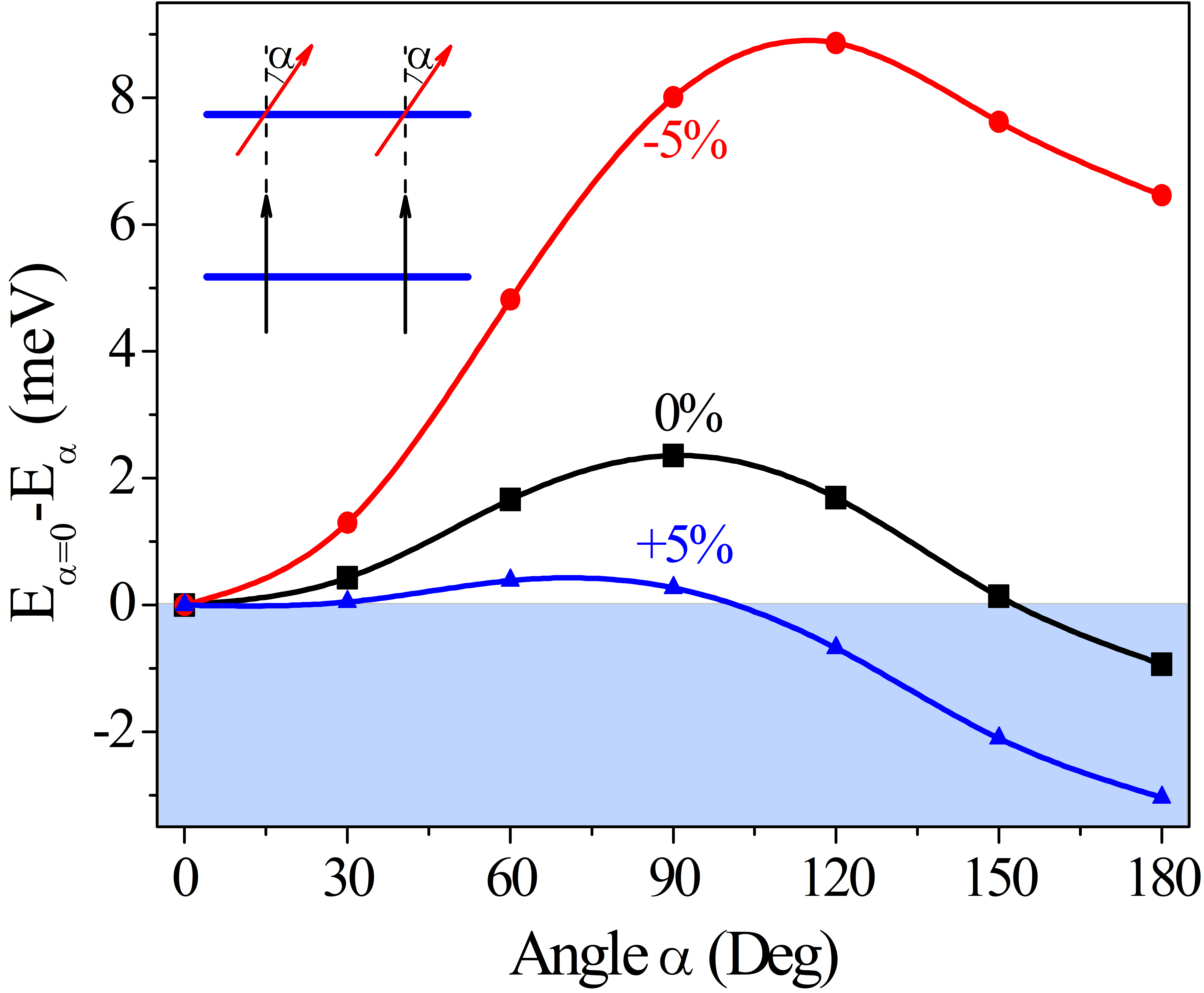} 
\caption{Total energy difference with respect to the FM state ($\alpha = 0$)
constraining the magnetization of one layer using LSDA+SOC.}
\label{Fig:mae}
\end{figure}

\subsection{\label{sec:no-mae} Layer Constrained Magnetic Anisotropy Energy} 
To check the reliability of our noncollinear calculations using LSDA+SOC in OpenMx, we have 
calculated the magnetic anisotropy energy (MAE) for the monolayer using the standard definition 
$\varepsilon_{MAE} = E_{\perp}-E_{//}$. For 0\% strain, we 
obtain a MAE of $\varepsilon_{MAE} = 0.68$ meV per Cr atom with an easy axis 
perpendicular to the atomic plane, consistent with the
previously reported value $\varepsilon_{MAE} = 0.65$ meV for all-electron 
calculations\cite{lado2017}, and $\varepsilon_{MAE} =0.98$ meV for a self-consistent 
approach\cite{jiang2018spin}. 
If we apply strain we find a maximum of the MAE for a -6\% strain, $\varepsilon_{MAE} = 1.5$ meV, however when we apply a tensile strain the value decreases reaching a value of $\varepsilon_{MAE} = 0.32$ meV per Cr atom for a 6\% strain. 

As we comment before, several magnetic ordering are possible in the 
bilayer structure, it is not possible to unequivocally define a magnetic
anisotropy energy valid in all cases. 
To get a sense of the variations in energy associated with the change in 
the direction of the magnetic moment, we focus first on a configuration 
where the Cr atoms of each layer are ferromagnetically coupled. 
Then, we fix the magnetic moment of the lower layer in the positive 
z-direction while we change the direction of the magnetic moment 
in the upper layer. We call it Layer Constrained Magnetic Anisotropy Energy (LC-MAE).
In the fig. \ref{Fig:mae}, the angle 0$^\circ$ corresponds to a ferromagnetic coupling 
between layers ($\upuparrows/\upuparrows$), and the angle of 180$^\circ$ corresponds to 
an antiferromagnetic coupling between layers ($\upuparrows/\downdownarrows$). 
We set the zero energy for each curve to the magnetic FM  ordering
($\upuparrows/\upuparrows$), where all spins are aligned ($\alpha=$0$^\circ$).
A lower 
value will mean a more stable phase.

Another way to read this graph would be that the higher the anisotropy value the more 
difficult to achieve that particular configuration, that way the higher 
values are obtained for $\alpha\approx $90$^\circ$-120$^\circ$. 
When $\alpha$=180$^\circ$ it can be seen that without and for a tensile 
strain the most stable configuration is AF while for a compressive strain of 
5\% to change to an AF  ($\upuparrows/\downdownarrows$) configuration is quite 
expensive.  


\section{Final Remarks}

We found that strain can be an efficient way to control the magnetic
properties of bilayer CrI$_3$. 
A tensile strain can stabilize the AF phase of the HT bilayer system while the compression causes the magnetic phase change from AF to FM, the higher the compression, the higher the potential barrier to be overcome to return to the AF phase. 
We showed that the magnetic phase change is related to an induced horizontal shift of the layers, our results are in agreement with recent experimental measurements\cite{Li2019,Song2019}. 
The small amount of strain required to modify the magnetic coupling between layers may be produced by lattice mismatch with the substrate and other interactions with the surrounding material --substrate, contacts, or encapsulating material--. Our work can drive to design functional devices combining magnetic and mechanical properties based on layered CrI3, and the reported phase diagram in CrI$_3$ mono- and bi-layers may help researchers to propose new heterostructures based on CrI$_3$ and its relatives.

\begin{acknowledgments}
AL acknowledges the financial support of CONICYT Becas Chile Postdoctorado Grant No. 74190099.
JWG acknowledges financial support from FONDECYT: Iniciaci\'on en Investigaci\'on 2019 grant N. 11190934 (Chile).
ESM acknowledges financial support from FONDECYT Regular 1170921 (Chile). 
JML acknowledges support from Financiamiento basal para centros cient\'ificos y tecnol\'ogicos de excelencia FB 0807.
Powered@NLHPC: This research was partially supported by the supercomputing infrastructure of the NLHPC (ECM-02).
The authors would like to thank F. Torres and F. Delgado for helpful discussions.
\end{acknowledgments}


\pagebreak
\newpage
\widetext

\begin{center}
\textbf{\large Supplemental Material: Strain-induced phase transition in CrI$_{3}$ bilayers}

Andrea Leon,$^1$ J. W. Gonz\'alez,$^2$, J. Mej\'ia-L\'opez,$^3$ F. Crasto de Lima,$^4$ and E. Suárez Morell.$^2$

$^1$ Max Planck Institute for Chemical Physics of Solids, Dresden, Germany.\\
$^2$ Departamento de F\'isica, Universidad T\'ecnica Federico Santa Mar\'ia, Casilla Postal 110V, Valpara\'iso, Chile.\\
$^3$ Centro de Investigaci\'on en Nanotecnolog\'ia y Materiales Avanzados CIEN-UC,
Facultad de F\'isica, Pontificia Universidad Cat\'olica de Chile, CEDENNA, Santiago.\\
$^4$ Instituto de F\'isica, Universidade Federal de Uberlândia, Uberlândia, MG, Brazil.

\end{center}

\setcounter{figure}{0} 
\setcounter{section}{0} 
\setcounter{equation}{0}
\setcounter{page}{1}
\renewcommand{\thepage}{S\arabic{page}} 
\renewcommand{\thesection}{S\Roman{section}}   
\renewcommand{\thetable}{S\arabic{table}}  
\renewcommand{\thefigure}{S\arabic{figure}} 
\renewcommand{\theequation}{S\arabic{equation}} 

\section{Methodology}
\textbf{OpenMX} uses a linear combination of pseudo-atomic orbitals (PAO). After a convergence check, we set a s3p3d2/s3p3d2f1 basis set and a cutoff radius of 6.0/7.0 a.u. for Cr/I atoms in our calculations\cite{liu2018electrical,han2004electronic} .
We solve the fully relativistic Kohn-Sham equation\cite{macdonald1979relativistic} 
in the LSDA and the PBE 
exchange-correlation functional are used with a plane-wave cutoff of $300$ Ry. 
The strong interactions between the 3d electrons in the Cr atoms are taking into account by adding the Hubbard parameter U. We used a typical value of $U = 3$ eV in the dual occupation approach\cite{han2007magnetic}.
All calculations, including the one to obtain the exchange coupling parameter J, were done with a  dense $15 \times 15 \times 1$ k-mesh and we considered a D2 method for the vdW  correction\cite{grimme2006semiempirical}.  
The procedure for all different strains consists in a modification of the lattice constant and then a relaxation of the atomic positions keeping the lattice constants fixed.

\textbf{VASP} calculations were performed using projected augmented wave (PAW) potentials 
to describe the core electrons as implemented in the Vienna ab-initio Simulation Package 
(VASP)\cite{VASP} and the generalized gradient approximation of Perdew-Burke-Ernzerhof 
(PBE) to account for the exchange-correlation energy of interacting electrons. 
The Brillouin zone was sampled with a 10$\times$10$\times$1 k-point mesh, a  plane wave 
basis with a kinetic energy cutoff of 500 eV was used. To avoid spurious interlayer 
interaction between periodic structures a 20 \AA{}   spacing of vacuum was introduced along the z-axis. 
Structural optimization were performed by the conjugate-gradient scheme until the  
Hellmann-Feynman force on each atom becomes smaller than 2 meV/\AA{}  and the total energy 
was converged to be within 10$^{-8}$ eV.
The calculation were performed taken into account Van der Waals interaction in the optB86b-vdW approximation\cite{klimevs2009chemical}.
A on-site Hubbard U Coulomb parameter of 3 eV was chosen  to account for Cr $d$-orbital strong electronic correlations.

\section{\label{sec:Disp} Displacement-induced phase transitions} 
\begin{figure}[!ht]
\centering
\includegraphics[clip,width=0.3\textwidth,angle=0]{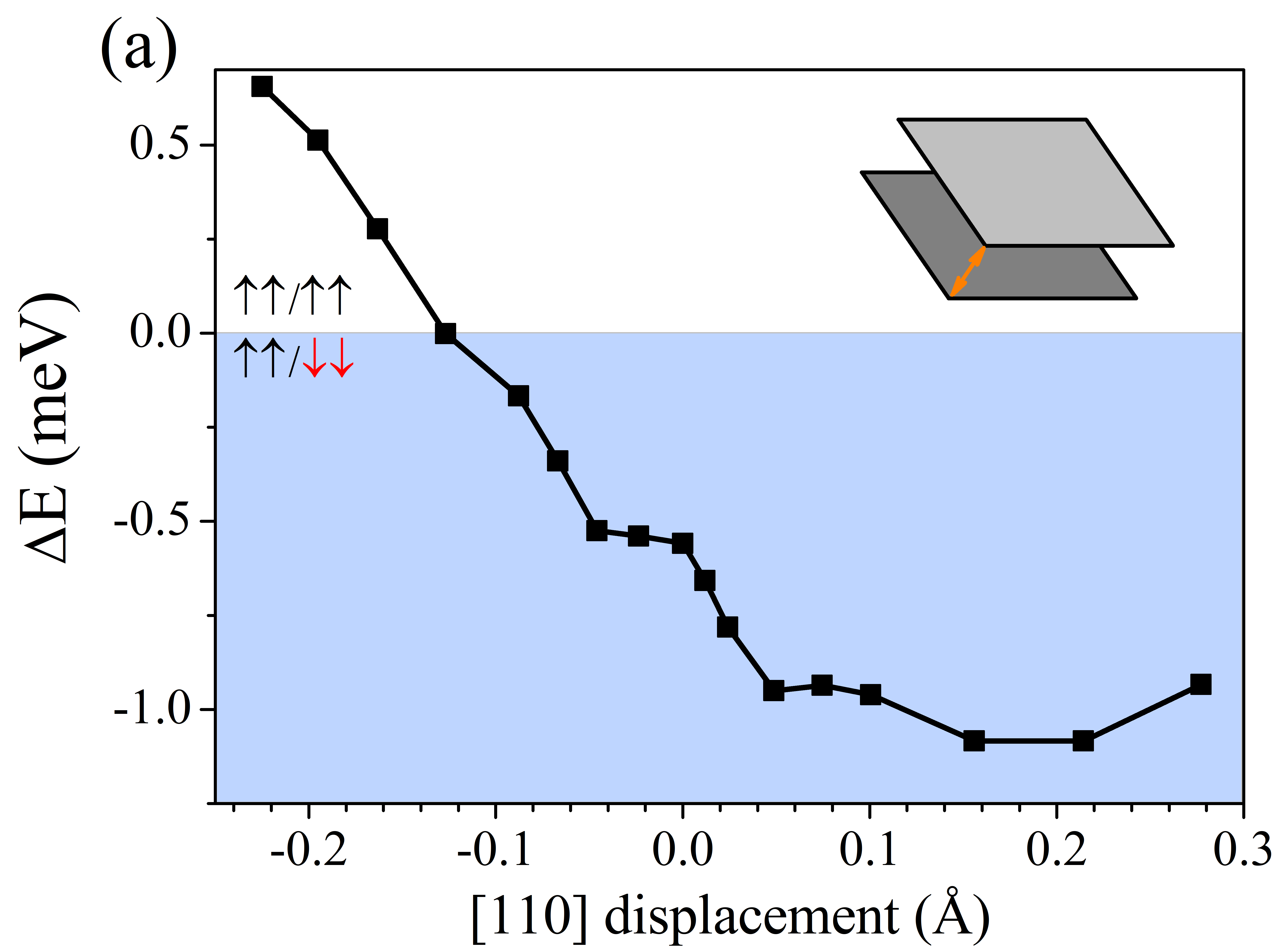} 
\includegraphics[clip,width=0.3\textwidth,angle=0]{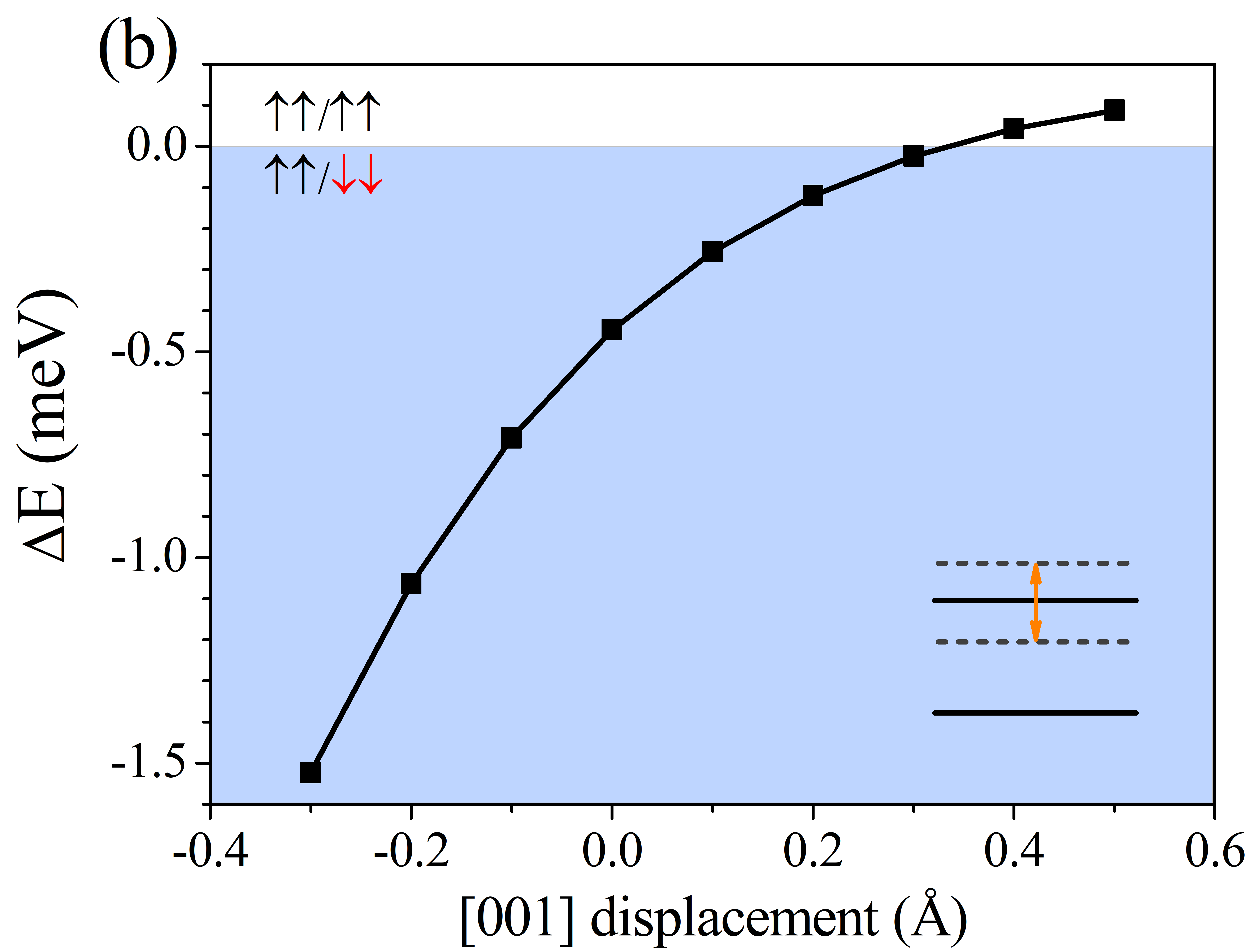} 
\caption{Energy change as a function of displacement between layers for 0\% strain. The  in-plane displacement in the direction [110] in (a) and the out-plane displacement [001] in (b).}
\label{Fig:dxyz_bi}
\end{figure}

In the main text, we showed that small displacements in the atoms position induced by the strain are responsible for the change of the magnetic coupling between layers that leads to a magnetic phase change. We have calculated how a shift of one of the layers along direction [110] in fig. \ref{Fig:dxyz_bi}(a) and in the out-plane direction [001] in fig. \ref{Fig:dxyz_bi}(b) changes the magnetic phase in both cases without strain. 

We observe a change in the sign of the magnetic coupling between layers by an small horizontal displacement of -0.1 \AA{}. This value is close to the one we obtained when the structure is compressed. If we change the distance between layers, we need to separate the layers by 0.4 \AA{} to obtain a transition from AF to FM, the system behaves as two separate layers, no changes are found when we compress vertically the structure up to a value of 0.3 \AA{}.

\section{\label{sec:Bands} Band structure}
\begin{figure}[!ht]
\centering
\includegraphics[clip,width=0.58\textwidth,angle=0]{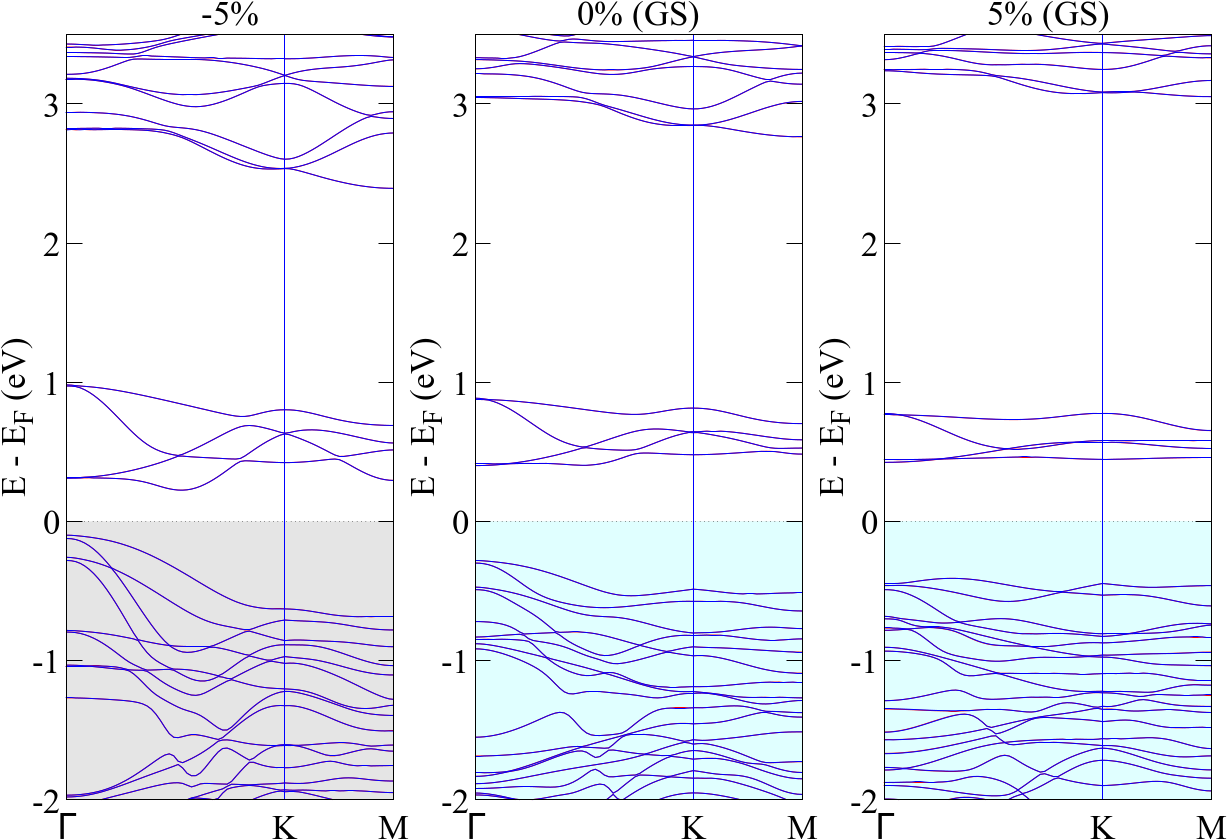} 
\includegraphics[clip,width=0.58\textwidth,angle=0]{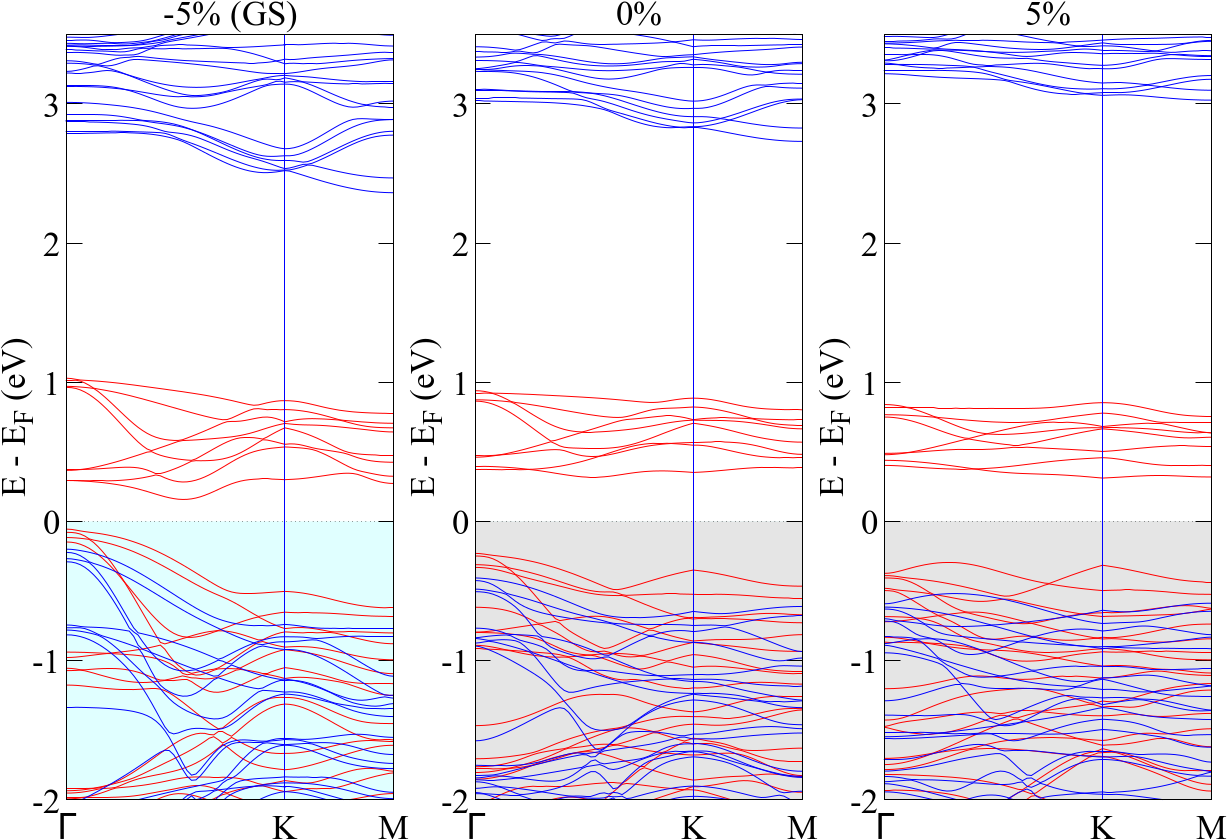} 
\caption{Band structure of the CrI$_3$ bilayer, in the top panel for AF-coupled layers 
($\upuparrows/\downdownarrows$) and in the bottom panel for the FM-coupled layers ($\upuparrows/\upuparrows$) 
in bottom panel. Red and blue lines correspond to spin up and spin down components. The ground state between the two phases at a given strain is marked as GS next to the value of the strain}
\label{Fig:bands_bi}
\end{figure}

In fig. \ref{Fig:bands_bi}, we show the band structure of the bilayer system calculated using PBE+U+vdW for different strain values. The top panel shows the evolution of the antiferromagnetic coupled layers ($\upuparrows/\downdownarrows$) and the bottom panel shows the ferromagnetic coupled layers ($\upuparrows/\upuparrows$).

The band structure of the CrI$_3$ bilayer (fig. \ref{Fig:bands_bi}) reveals that the bilayer HT is a semiconductor with a direct bandgap of 0.68 eV for the AF-coupled layers (ground state) and the FM-coupled bilayer is a semiconductor with an indirect bandgap of 0.53 eV. Both results are in agreement with previous reports\cite{jiang2019stacking}.
In the AF-coupled layers ($\upuparrows/\downdownarrows$) the top of the valence band appears in the $\Gamma$-point at -0.1 eV for -5\% strain and -0.3 eV for 0\%, on the other hand, for 5\% the highest occupied state is halfway between $\Gamma$ and $K$ points. 
A similar scenario is repeated for the FM-coupled layers ($\upuparrows/\upuparrows$). In general, the spin up bands are slightly higher in energy than the spin down bands. At the $\Gamma$-point, for -5\% strain the top of the spin up (down) band appears at -0.05 eV (-0.2 eV) and for 0\% the top is at -0.23 eV (-0.4 eV). For a 5\% positive strain, the top of the band is -0.3 eV (-0.5 eV). 

.

%

\section{\label{sec:Ising} Ising model} 
The Ising model is widely used to describe the magnetic coupling of low-dimensional magnetic 
systems\cite{chen2019boosting,miao20182d,lu2019mechanical}. Herein, we numerically solve the 
Ising model to calculate the Curie temperature T$_{C}$. 
In our simplified model, we include  nearest-neighboring exchange interactions in mono- and 
bi-layers. The spin Hamiltonian defined on the honeycomb lattice is
\begin{equation}
\mathcal{H} = - \sum_{\langle i,j \rangle} J_{ij} S_i \cdot S_j,
\end{equation}
where the summation $\langle i,j \rangle$ runs over nearest-neighbor Cr sites 
(3 in-plane first-neighbors, 2 out-plane first-neighbors and 1 out-plane second-neighbor),
J$_{ij}$ is the exchange parameter calculated in a previous section 
(see fig. \ref{Fig:Jbi} and related comments) and 
$S_{i \, (j)}$ is the spin $S = \pm 3/2$ of Cr atoms. 

Using Metropolis Monte Carlo simulations based on this 2D Ising 
model,\cite{paez2007computational} we have estimated the Curie temperature and its 
response upon strain in CrI$_{3}$ mono- and bi-layers. 
After a convergence test, a $20 \times 20$ hexagonal superlattice with periodic
boundary condition are used. The thermodynamic properties are extracted after the system 
reaches the equilibrium state for a given temperature. The specific heat can be determine 
from the energy fluctuations:
\begin{equation}
C_V = \frac{1}{N} \frac{dE}{dT}= \frac{1}{N^2}\frac{\langle E^2 \rangle - \langle E \rangle ^2}
{k_B T^2},
\end{equation}
where $E$ is the internal energy, $N$ is the number of magnetic sites 
and $k_B$  Boltzmann's constant. 

At a critical temperature, the system undergoes a 2nd order phase transition, 
for temperatures below the Curie temperature the effect of the exchange $J$ is 
large enough to produce spontaneous alignment of the neighbouring atomic spins.
The Curie temperature, T$_C$  is extracted from the peak of the specific heat. In particular, 
we obtain the T$_C$ by fitting the numerical solution data from the Ising model to a Lorentzian 
function. 

\begin{figure}[!hb]
\centering
\includegraphics[clip,width=0.45\textwidth,angle=0]{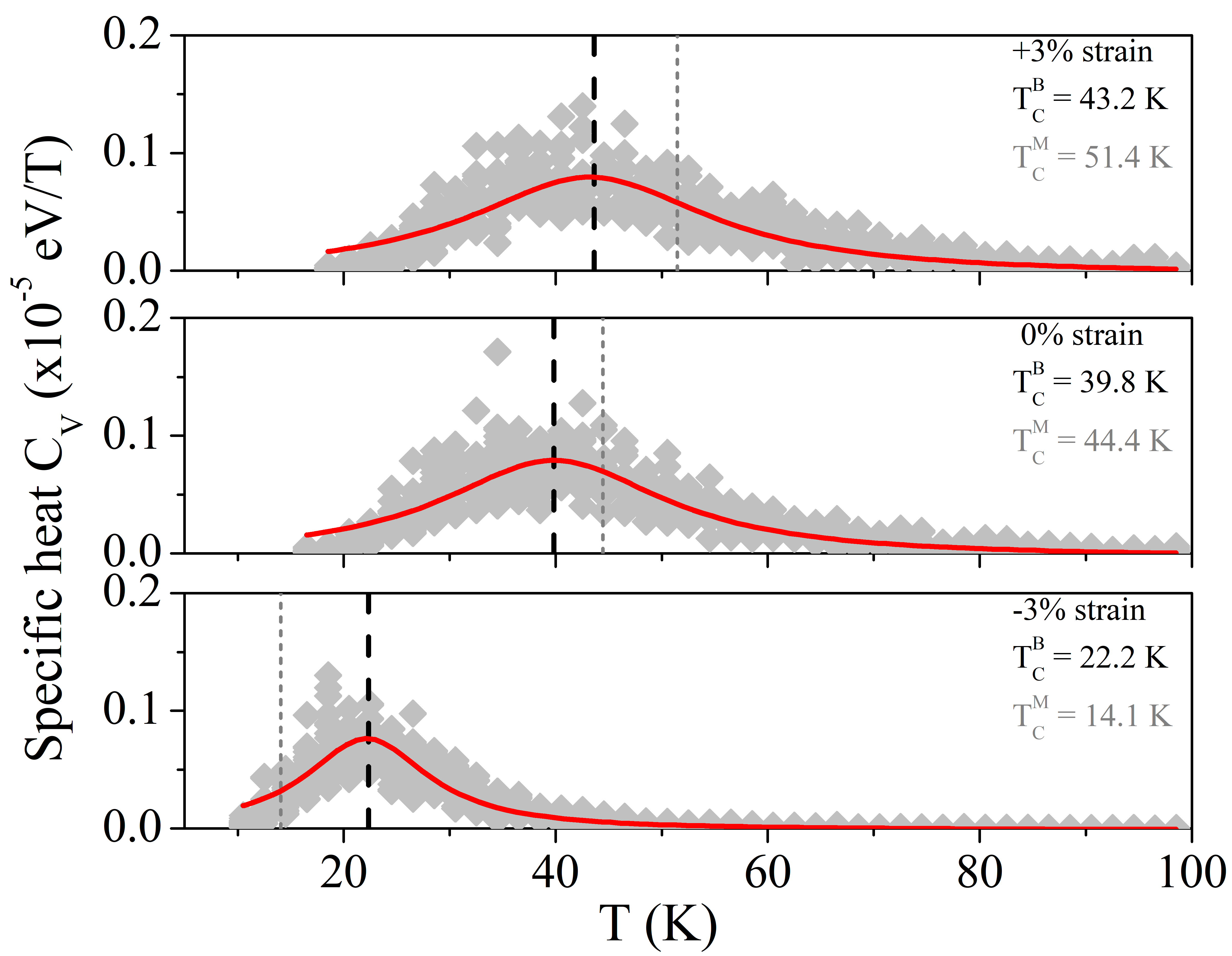} 
\caption{Specific heat C$_{V}$ for the bilayer as a function of the temperature for +3\%, 0\% and -3\% strain. 
The Curie temperature (T$^B_C$ indicated with a black vertical line) 
and the fitting of data from the Ising model to a Lorentzian function shown in red. 
For a comparison, the Curie temperature of the monolayer T$^M_C$ is indicated as a grey line. }
\label{Fig:isi_bi}
\end{figure}

\section{VASP Results}

To validate the results discussed in the main text, we present a summary with  the main findings using VASP.

\subsection{\label{mono} Monolayer}

\begin{figure}[ht!]
\centering
\includegraphics[clip,width=0.65\textwidth,angle=0]{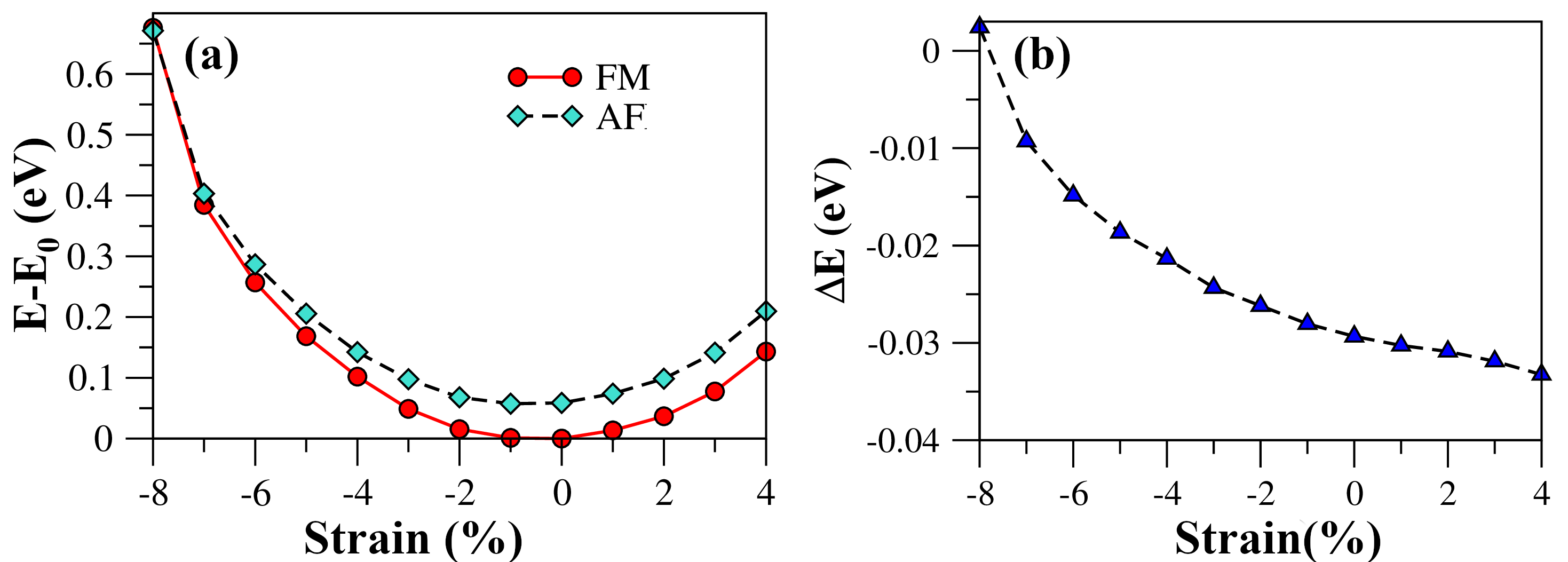} 
\caption{Monolayer VASP calculations. (a) Energy difference E-E$_{0}$ as a function of strain 
for the two configurations FM and AF, where E$_{0}$ corresponds to the minimun energy of the FM configuration. In (b) the energy difference $\Delta$E = E$_{FM}$-E$_{AF}$ 
per Cr atoms as a function of the strain.}
\label{Fig:vasp_mono}
\end{figure}

In fig. \ref{Fig:vasp_mono}, we show the results of the calculation for the two magnetic phases FM and AF for the bilayer HT structure. This is similar to fig. \ref{Fig:Energy_mono} in the main text and the results are in excellent agreement. In this case we set the zero energy to the value of the total energy without strain of the FM phase.

\subsection{\label{bi} Bilayer}
\begin{figure}[hb!]
\centering
\includegraphics[clip,width=0.80\textwidth,angle=0]{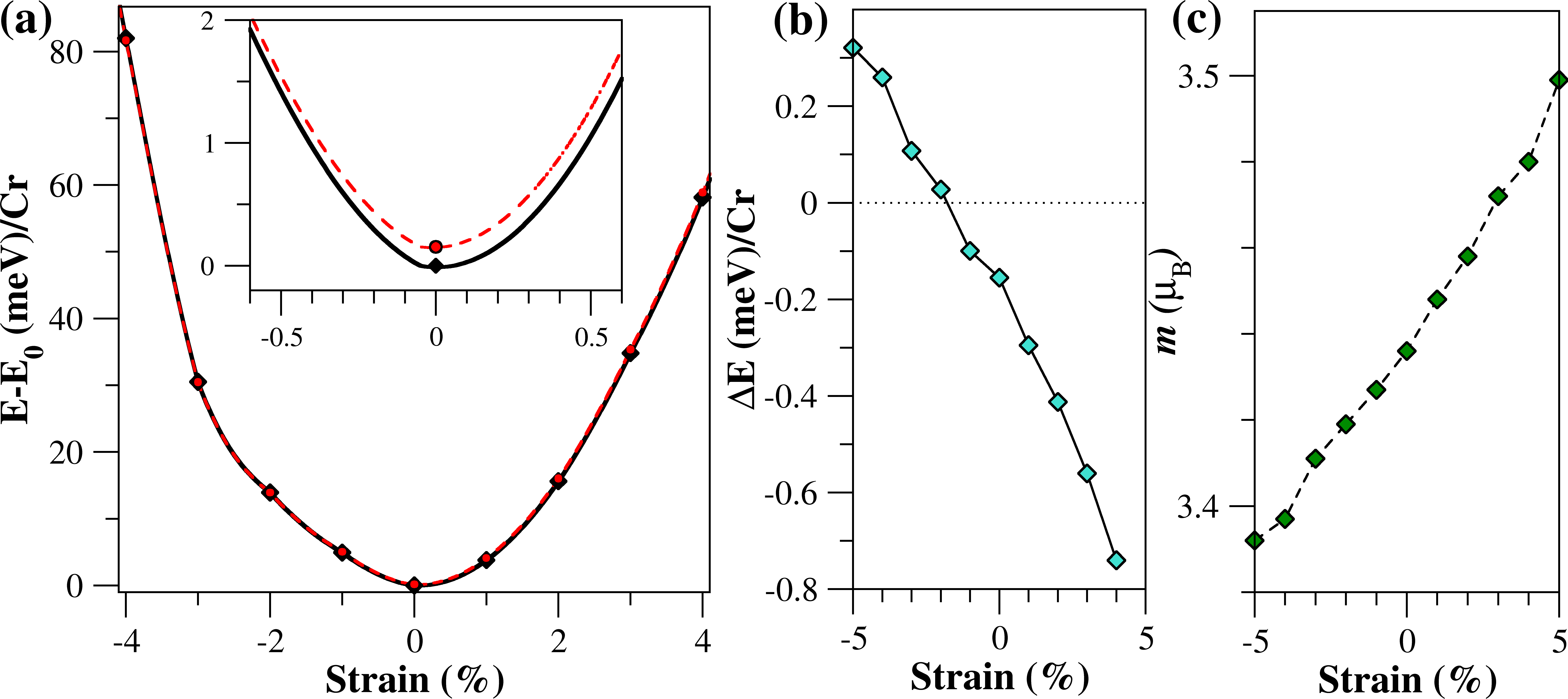} 
\caption{Bilayer VASP calculations.
In (a) Energy (E-E$_{0}$) as a function of the strain for FM and AF configuration (E$_{0}$ corresponds to the energy minimum for AF configuration). Inset: Zoom around 0\% strain.
In (b) energy difference $\Delta$E per Cr atom 
($\Delta$E = $E_{\upuparrows/\upuparrows}-E_{\upuparrows/\downdownarrows}$) as a function of the strain. In (c) the average value of absolute magnetic moment m per Cr atom. }
\label{Fig:vasp_bilayer}
\end{figure}

We have calculated using VASP the total energy of each configuration as a function of the strain, the results are in fig. \ref{Fig:vasp_bilayer}a,  In fig. \ref{Fig:vasp_bilayer}d we show a zoom around 0\% strain of the same results. The difference in energy between the two phases is plotted in \ref{Fig:vasp_bilayer}b. The transition from AF to FM is around -2\% (compressive strain). We also calculated how the magnetic moment of Cr atoms changes with the strain. The differences are small, only with variations of 0.1 $\mu_B$ along the whole range studied.

\section{Comparison between methods\label{compa}}
In this section, we demonstrate the reliability of our conclusions comparing with other approaches. In table \ref{table_omx} we summarize the OpenMX results and in table \ref{table_vasp} the VASP results. 
For strain 0\% and -5\%, we report the lattice constant, the out of plane separation between Cr atoms of different layers and the $E_{\upuparrows/\upuparrows}-E_{\upuparrows/\downdownarrows}$ energy difference  (positive means $\upuparrows/\downdownarrows$ stability). 
We find that the critical compression for a change in the sign of the interlayer coupling is around -3\% (with a variations around $\pm$ 1) depending on the parameters and internal details of the selected approach.
Note that for all approaches, the ground state in the no-strain case (0\%) is $\upuparrows/\downdownarrows$, and for compression of -5\%, the $\upuparrows/\upuparrows$ configuration is the ground state. 

Our OpenMX results using an empirical damped dispersion correction (see table \ref{table_omx}) are in agreement with previous DFT calculations at PBE and LDA level\cite{jiang2019stacking,liu2018electrical} and with our non-local vdW corrections with VASP. As expected, when we compare with experimental data, the LDA underestimates the lattice 
constant, while PBE overestimates it. The lattice constant identified using X-ray diffraction and absorption spectroscopy techniques in bulk/few-layers\cite{frisk2018magnetic,Huang2017,mcguire2015coupling} 
CrI$_3$ is 6.87 \AA{}. 

\begin{table}[ht]
\centering
\begin{tabular}{||c||c|c|c||c|c|c||}
\hline 
 & \multicolumn{3}{c||}{$0\%$  strain} & \multicolumn{3}{c||}{$-5\%$ strain} \\ 
\hline 
Approach & a (\AA) & d$_z$ (\AA) & $\Delta E$ (meV) & a (\AA) & d$_z$ (\AA) & $\Delta E$ (meV) \\ 
\hline 
\hline 
LDA+vdWD2 & 6.70 & 6.34 & 0.4 & 6.36 & 6.53 & -0.1 \\ 
\hline 
LDA+vdWD3 & 6.77 & 6.42 & 0.4 & 6.43 & 6.63 & -0.1 \\ 
\hline 
PBE+vdWD2 & 6.94 & 6.59 & 0.6 & 6.60 & 6.78 & -0.2 \\ 
\hline 
PBE+vdWD3 & 7.09 & 7.16 & 0.5 & 6.74 & 7.37 & -0.1 \\ 
\hline 
\end{tabular}
\caption{Summary of the OpenMX results with an empirical damped dispersion correction, to account to the vdW interactions, for the ground state ($0\%$)  and strained ($-5\%$) systems, where a, d$_z$ and $\Delta E$ indicates the lattice constant, interlayer Cr-Cr distance, and energy difference between $\upuparrows/\upuparrows$ and $\upuparrows/\downdownarrows$ phases. Positive (negative) $\Delta E$ indicate the stability of $\upuparrows/\downdownarrows$ ($\upuparrows/\upuparrows$).}
\label{table_omx}
\end{table}

The VASP results within the non-local vdW approaches considering the vdW-DF,
proposed by Dion et. al. \cite{PRLdion2004}, and its variants, namely the vdW-DF2 \cite{PRBlee2010}, optB86b-vdW \cite{JPCMklime2009}, and the rev-vdW-DF2 \cite{PRBhamada2014}, with its results summarized in Table~\ref{table_vasp}. We observe an overestimation of the lattice parameter in relation to the experimental measurements (6.87\,{\AA}), however the optB86b-vdW and rev-vdW-DF2 approach deviates by less than $1\%$, given its good description of layered materials \cite{PRMtran2019}. Nevertheless, all the explored functionals corroborate our conclusions describing the magnetic $\upuparrows/\downdownarrows$ ground state ($0\%$), while still predicting the magnetic phase transition, for instance, to $\upuparrows/\upuparrows$ upon $-5\%$ strain ruled by the increased distance between the layers. For instance, for rev-vdW-DF2 the difference in energy $\Delta E$ changes from  $\Delta E= 0.4 \rightarrow -0.9$\,eV as the system is strained, from $0\%$ to $-5\%$. Besides the effect of the vdW functionals to the atomic positions, it also contribute to the electronic structure. \\
In order to isolate the effect of the non-local vdW dispersion to the magnetic phase alone, we have performed PBE calculations keeping the system constrained to the vdW optimized geometries, to which $\Delta E$ are shown in parenthesis on Table~\ref{table_vasp}. Here, we can see that besides the atomic structure changes, the non-local interactions tends to further stabilize the $\upuparrows/\downdownarrows$ phase, which are seem by the higher $\Delta E$ compared to the constrained PBE approach.


\begin{table}[ht]
\centering
\begin{tabular}{||c||c|c|c||c|c|c||}
\hline 
 & \multicolumn{3}{c||}{$0\%$  strain} & \multicolumn{3}{c||}{$-5\%$ strain} \\ 
\hline 
Approach & a (\AA) & d$_z$ (\AA) & $\Delta E$ (meV) & a (\AA) & d$_z$ (\AA) & $\Delta E$ (meV) \\ 
\hline 
\hline 
vdW-DF      & 7.19 & 7.09 & 0.4 (0.2) & 6.83 & 7.25 & -0.5 (-0.9) \\ 
\hline 
vdW-DF2     & 7.21 & 7.00 & 0.3 (0.3) & 6.85 & 7.18 & -0.7 (-1.1) \\ 
\hline
optB86b-vdW & 6.93 & 6.67 & 0.3 (0.0) & 6.58 & 6.82 & -1.3 (-1.4) \\
\hline
rev-vdW-DF2 & 6.93 & 6.68 & 0.4 (0.1) & 6.58 & 6.83 & -0.9 (-1.4) \\ 
\hline 
\end{tabular}
\caption{Summary of the VASP results with a non-local dispersion correction, to account to the vdW interactions, for the ground state ($0\%$)  and strained ($-5\%$) systems, where a, d$_z$ and $\Delta E$ indicates the lattice constant, interlayer Cr-Cr distance, and energy difference between $\upuparrows/\upuparrows$ and $\upuparrows/\downdownarrows$ phases. Positive (negative) $\Delta E$ indicate the stability of $\upuparrows/\downdownarrows$ ($\upuparrows/\upuparrows$). In parenthesis we show the $\Delta E$ for a constrained PBE calculation keeping the vdW atomic structure.}
\label{table_vasp}
\end{table}


\end{document}